\documentclass[pre,showpacs,twocolumn,superscriptaddress]{revtex4}
\usepackage{amssymb,graphicx,amsbsy}
\usepackage{color}
\usepackage{bm}

\begin{document}

\title{Phase transition of $q$-state clock models on heptagonal lattices}
\author{Seung Ki Baek}
\affiliation{Department of Physics, Ume{\aa} University, 901 87 Ume{\aa},
Sweden}
\author{Petter Minnhagen}
\affiliation{Department of Physics, Ume{\aa} University, 901 87 Ume{\aa},
Sweden}
\author{Hiroyuki Shima}
\affiliation{Department of Applied Physics, Graduate School of Engineering,
Hokkaido University, Sapporo 060-8628, Japan}
\author{Beom Jun Kim}
\email[Corresponding author, E-mail: ]{beomjun@skku.edu}
\affiliation{BK21 Physics Research Division and Department of Energy Science,
Sungkyunkwan University, Suwon 440-746, Korea}
\affiliation{Department of Computational Biology, School of Computer Science
and Communication, Royal Institute of Technology, 100 44 Stockholm, Sweden}

\begin{abstract}
We study the $q$-state clock models on heptagonal lattices
assigned on a negatively curved surface.
We show that
the system exhibits three classes of equilibrium phases;
in between ordered and disordered phases, 
an intermediate phase 
characterized by a diverging susceptibility 
with no magnetic order
is observed at every $q \ge 2$.
The persistence of the third phase for all $q$
is in contrast with the disappearance of the 
counterpart phase
in a planar system for small $q$,
which indicates the significance of 
nonvanishing surface-volume ratio 
that is peculiar in the heptagonal lattice.
Analytic arguments 
based on Ginzburg-Landau theory and generalized Cayley trees
make clear that the two-stage transition in the present system is attributed to
an energy gap of spin-wave excitations
and strong boundary-spin contributions.
We further demonstrate that boundary effects breaks 
the mean-field character in the bulk region,
which establishes the consistency with results of clock models on
boundary-free hyperbolic lattices.
\end{abstract}


\maketitle

\section{Introduction}

The role of geometry has continued  drawing attention
in statistical physics. A curved surface, for example, has been a useful test
ground to study ergodicity~\cite{balazs}, and curved nanoscale
carbon structures have been expected to possess interesting elastic and
magnetic properties~\cite{nano}.
Recently, rapid development of soft material sciences also requires a
precise understanding of physics on a curved surface in terms of
geometric interactions~\cite{kamien-hemmen}.
One immediate question from the statistical-physical viewpoint is how
phase transitions occur on such a curved surface since they in general
depend on geometrical factors.
In particular, a negative Gaussian curvature yielding a saddle-like
hyperbolic surface
has been more commonly studied in critical phenomena than a positive one
since a positive curvature tends to make a closed surface so that it is hard
to extend the system size while keeping the magnitude of the curvature constant.
In a negatively curved surface, the length scale grows only logarithmically
with the 
surface area,
and thus one could expect a mean-field-like critical
behavior in many systems. Whereas this expectation was proven true for
the bulk of
the Ising spin system~\cite{shima,ueda}, the $XY$ spin model has no local order
at finite temperatures~\cite{xy}. This lack of order
in the $XY$ model
is attributed to the 
gapless
spin-wave excitations
that can arise from the boundary
at any finite temperature $T$.
This argument is based on the fact that a negatively curved surface contains
a huge amount of boundary points: that is, 
for a negatively curved surface,
the ratio of surface area to perimeter (which is the two-dimensional example
of the so-called surface-volume ratio in general dimension) remain
nonvanishing even in the large-system limit.
Since it was pointed out that a system may have a novel
behavior due to the presence of a nonvanishing boundary~\cite{auriac},
there have been ongoing
studies to clarify this issue~\cite{shima,shima2006a,xy,diff,perc,geomxy}.
While the boundary effects can be sometimes excluded, for example, by using a
periodic boundary condition~\cite{sausset} or by mathematical
abstractions~\cite{belo,ueda,gendiar,krcmar}, it is often crucial to
understand how a boundary affects the physical properties
since it may give the most important contribution to an observed behavior as
will be explained in this work.

The complete difference between the Ising and the $XY$ models with respect to
the presence or absence of the ordered phase motivates us to study the
$q$-state clock model
on a negatively curved surface.
The $q$-state clock model
is equivalent to the Ising model for $q=2$ and
approaches to the $XY$ model for $q \rightarrow \infty$. Thereby one can
obtain a better understanding on how the phase structure changes in between
with varying $q$.
In this paper, we present the following findings: first, the critical
temperature $T_c$ is indeed proportional to the energy gap to excite the
spin fluctuations.
second, we report an intermediate phase with a diverging susceptibility
between the ordered and disordered phases. While it corresponds to the
quasiliquid phase in the planar case, an interesting difference is that
this intermediate region characterized by the vanishing order parameter and
diverging susceptibility
is observable at every $q \ge 2$ on the curved
structure. This point will be further discussed by studying the Cayley tree
analytically.

This work is organized as follows: in Sec.~\ref{sec:hyperb}, we explain the
construction of our lattice for describing a negatively curved surface, and
introduce the $q$-state clock model on top of it. The
results will be presented and discussed in Sec.~\ref{sec:result}.
We then summarize this work in Sec.~\ref{sec:summary}.

\section{Clock model in hyperbolic lattice}
\label{sec:hyperb}

\begin{figure}
\includegraphics[width=0.2\textwidth]{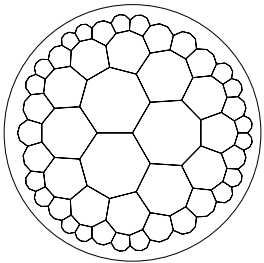}
\caption{Schematic view of a heptagonal lattice with a level $l=3$,
projected on the Poincar\'e disk.}
\label{fig:hyperb}
\end{figure}

A Schl\"afli symbol $\{k,w\}$ means a tessellation that $w$ regular
$k$-sided polygons meet at each vertex.
Satisfying $(k-2)(w-2) > 4$,
every pair of $\{k,w\}$ results in a negatively curved
surface, yielding a hyperbolic tessellation~\cite{coxeter1997}.
Each hyperbolic tessellation gives a resulting lattice structure,
which will be generally called a hyperbolic lattice.
In this work, we construct one type of hyperbolic lattices, i.e.,
a heptagonal lattice denoted as $\{k,w\} =
\{7,3\}$, in a concentric way as depicted in Fig.~\ref{fig:hyperb}.
We start with the zeroth layer, a
point in the middle of the Poincar\'e disk~\cite{green},
and surround it by three
heptagons. Then the newly added 15 points constitute the first layer.
Likewise, attaching 12 heptagons all the way around the first layer adds 45
more points, which make the second layer, and so on. A heptagonal lattice
of a level $l$ means that it is made up to the $l$th layer, and its
system size is then given by $N(l) = 1 + \frac{15}{\sqrt{5}} \sum_{j=1}^{l}
[(\frac{3 + \sqrt{5}}{2})^j - (\frac{3 - \sqrt{5}}{2})^j]$. As $N(l)$
increases exponentially with $l$, the surface-volume ratio does not
vanish even in the large-size limit. 

An important consequence of the non-vanishing surface-volume ratio
is an enhancement of boundary effects
that exceeds the bulk-spin contributions.
Sometimes only the bulk properties
are studied by restricting ourselves to a distance less than $x l$ from
the zeroth layer with a constant $0<x<1$. However, one should remember that
the system would not be properly
described by the bulk part since its fraction eventually vanishes:
suppose that $N(l) \sim e^{zl}$ for some curvature-dependent constant $z$.
The bulk fraction is
then $e^{xzl}/e^{zl} = e^{-(1-x)zl}$, which exponentially decreases as
$l$ grows.
This is why the boundary-spin contribution plays a dominant role
in determining the physical properties of the whole system.

By the $q$-state clock model, we mean a spin system described by the
following Hamiltonian:
\begin{eqnarray}
H &=& -J \sum_{\left< ij \right>} \boldsymbol{s}_i \cdot \boldsymbol{s}_j -
\sum_i \boldsymbol{h} \cdot \boldsymbol{s}_i \nonumber\\
&=& -J \sum_{\left< ij \right>}
\cos(\theta_i - \theta_j) - \sum_i h \cos\theta_i,
\label{eq:hamiltonian}
\end{eqnarray}
where each spin $\boldsymbol{s}_i$ can have one of $q$ possible angles,
$\theta_i = 2 \pi n_i/q$ with $n_i = 0,1,\ldots,q-1$, and $\boldsymbol{h}$
is a magnetic field along the direction for $\theta=0$ with a magnitude $h$. 
The summation is over the nearest
neighbors, and the coupling constant $J>0$ is the strength of the
ferromagnetic interaction.
As mentioned
above, $q=2$ and $q=\infty$ correspond to the Ising and $XY$ models,
respectively. In addition, the case of $q=3$ is equivalent to the
three-state Potts
model~\cite{wu}. The case of $q=4$
has the same universality class as the Ising system
since the partition function of the four-state clock model at temperature
$T$ is formally isomorphic to that of two uncoupled Ising systems at
$T/2$~\cite{suzuki}.

In the planar case, the $q$-state clock model for $h=0$ generally has
three phases in the $q-T$ plane~\cite{lapilli}.
Two among the three are ordered and disordered phases as in the Ising model.
From the existence of the Kosterlitz-Thouless (KT)
phase in the $XY$ limit~\cite{bkt}, one can argue that
the third, quasiliquid phase emerges for $q > 4$ in the intermediate
temperature range~\cite{elitzur}. The low
transition point where the ordered phase vanishes is roughly described by
$T_1 \propto 1/q^2$, as explained in the Villain
approximation~\cite{elitzur,jkkn}. On the other hand, the high transition
point, where disordered phase begins, remains almost constant around $T_2 =
T_{\rm KT} \simeq 0.89 J/k_B$ for $q \ge 8$, where
$k_B$ is the Boltzmann constant~\cite{lapilli}.

With a constant negative curvature, as shown in the next section, some of
these behaviors still look qualitatively similar. Specifically, the lower
transition point is roughly proportional to $1/q^2$ whereas the higher one
does not change much as $q$ increases. However, there also exist clear
differences in that the intermediate phase between these two temperatures is
created by a very different mechanism discussed later, and is present at
every $q \ge 2$.

\section{Results}
\label{sec:result}

\subsection{Ginzburg-Landau theory for homogeneous lattice without boundary}
\label{sub:hom}

Phase transitions on a curved surface can be very different whether a boundary
of the system is considered or not. As our numerical experiments include
both of the curvature and boundary effects, we will first consider only the
curvature effects in this part, in order to highlight the boundary effects
more clearly.

Suppose the $q$-state clock model is in a continuum limit.
Phenomenologically one may write a dimensionless free energy $F$ of
this system in the ordered phase~\cite{free} as
\begin{widetext}
\begin{equation}
F = \int d \boldsymbol{\rho} \left[ |\nabla \psi|^2 - 
|\psi|^2 + \frac{1}{2} |\psi|^4 + \frac{v}{q} (\psi^q + {\psi^\ast}^q -2
|\psi|^q ) - \frac{1}{2} (\tilde{h} \psi^\ast + \tilde{h}^\ast \psi) \right],
\label{eq:free}
\end{equation}
\end{widetext}
where $\boldsymbol{\rho}$ is a position vector ($\boldsymbol{r}$) rescaled by 
a specific
length
scale,
$\xi$, so that $|\boldsymbol{\rho}| = |\boldsymbol{r}|/\xi$.
In Eq.~(\ref{eq:free}),
$\psi(\boldsymbol{\rho},t) = |\psi (\boldsymbol{\rho},t)|
\exp \left[i\phi(\bm{\rho},t) \right]$
is a complex order parameter,
$\tilde{h}(\boldsymbol{\rho}, t)$ is a dimensionless magnetic field
represented as a complex number,
and $v$ is a positive constant.
Functional differentiation of Eq.~(\ref{eq:free}) with respect
to $\psi^\ast$ yields
\begin{equation}
\frac{\delta F}{\delta \psi^\ast}
= -\nabla^2 \psi - \psi (1-|\psi|^2) + 
\frac{v ({\psi^\ast}^q -|\psi|^q)}{\psi^\ast}
- \tilde{h}.
\label{eq:motion}
\end{equation}
Assuming the free-energy minimum, $\delta F/\delta \psi^\ast = 0$,
we differentiate Eq.~(\ref{eq:motion})
with respect to
$\tilde{h}$ to find 
an
equation for the two-point correlation function,
$G(\boldsymbol{\rho}, \boldsymbol{\rho}') = \partial
\psi(\boldsymbol{\rho}) / \partial \tilde{h}(\boldsymbol{\rho}')$:
\begin{equation}
\left[ -\nabla^2 -1 + 2|\psi|^2 - \frac{qv}{2} |\psi|^{q-2} \right]
G(\boldsymbol{\rho}, \boldsymbol{\rho}') = \delta
(\boldsymbol{\rho},\boldsymbol{\rho}').
\end{equation}
For a translationally invariant system, we may set $\boldsymbol{\rho}'=0$
without loss of generality.
Let us take a sufficiently small $v$ 
so that this system has ground states with $|\psi|>0$~\cite{enomoto1990}.
Then one finds $|\psi| \approx 1$ for small $v$ from which it follows
\begin{equation}
( \nabla^2 - 1 ) G(\rho) = - \delta (\rho),
\label{eq:green}
\end{equation}
where $\rho \equiv |\boldsymbol{\rho}|$.

We now impose negative Gaussian curvature to the underlying surface of the model.
On a hyperbolic surface, the Laplacian operator is replaced by $\triangle$
written as~\cite{young}
\begin{eqnarray*}
\triangle &=& \frac{1}{\sinh \rho} \frac{\partial}{\partial
\rho} \left( \sinh \rho \frac{\partial}{\partial \rho} \right) +
\frac{1}{\sinh^2 \rho}
\frac{\partial^2}{\partial \theta^2}\\
&\approx& \frac{\partial^2}{\partial \rho^2}
+ \frac{\partial}{\partial \rho},
\end{eqnarray*}
where the approximation can be taken due to the exponential increase of
$\sinh \rho$. Then Eq.~(\ref{eq:green}) is reduced to
\begin{equation}
\left( \frac{\partial^2}{\partial \rho^2} + \frac{\partial}{\partial \rho} - 1
\right) G(\rho) = -\delta(\rho).
\end{equation}
This equation is solved yielding $ G(\rho) \sim
e^{-\rho(1+\sqrt{5})/2}$.
Note that the correlation function basically behaves like a
one-dimensional case
as
$G(r) \sim e^{-r/\xi}$~\cite{golden},
where $\xi$ serves as the correlation length of the system.
Such an exponential decay in $G(\rho)$ on the hyperbolic surface
apparently suggests the absence of order-disorder phase transition in the current system.
This is, however, not the case.
A noteworthy
point is that
the number of spins $N_s(r)$ within a
distance $r$ also increases exponentially in a hyperbolic surface,
which results in divergence of the magnetic susceptibility defined by
\begin{equation}
\chi = (N k_B T)^{-1} \sum_{i,j} G(i,j),
\label{eq:sum}
\end{equation}
where the summation is over every possible pair $(i,j)$ of spins.
Even if $G(r)$ shows an exponential decay,
susceptibility $\chi$ is able to diverge 
at finite $T$
by satisfying $N_s(r) G(r) \ge1$, 
from which the critical temperature $T_c$ can be located~\cite{lopes}.

In addition to the curvature effects mentioned above,
we should take note of the effects of strong boundary-spin contributions
that are inherent to the present system.
Notice that in passing from Eq.~(\ref{eq:free}) to Eq.~(\ref{eq:motion}),
we have discarded a surface term.
As mentioned already, 
however,
the boundary effect cannot be
neglected in any physically realizable system with a constant negative
curvature.
Hence, the present system involving the boundary effects will exhibit
distinct properties from the mean-field character observed in
Ref.~\cite{gendiar} wherein the boundary effects are artificially excluded.
We also note that our discussion in the previous paragraph supports the validity of
the mean-field description in the boundary-free system
since the correlation function decays so fast at $T_c$~\cite{golden}.

\subsection{Estimation of the Lower Transition Temperature $T_c$}
\label{sub:sim}

\begin{figure}
\includegraphics[width=0.45\textwidth]{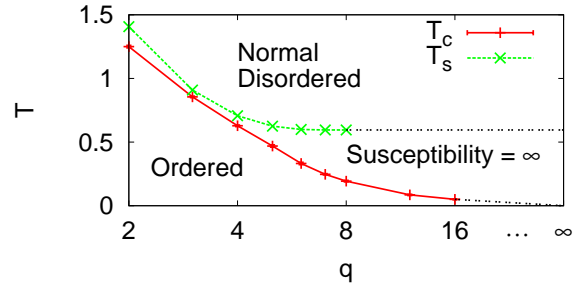}
\caption{(Color online) Phase diagram of the $q$-state clock model on the
heptagonal lattice as shown in Fig.~\ref{fig:hyperb}. We define two transition
temperatures $T_c$ and $T_s$ so that there exist the ordered phase below $T_c$
and the normal disordered phase above $T_s$. The intermediated phase
is characterized by a diverging susceptibility with no magnetic order.
The dotted lines mean extrapolated behaviors of
the transition temperatures to the $XY$-model limit ($q = \infty$).
}
\label{fig:phase}
\end{figure}

In Fig.~\ref{fig:phase}, we propose a phase diagram of the clock model on
the physically realizable hyperbolic lattice introduced in
Sec.~\ref{sec:hyperb}.
In this diagram, we define $T_c$ as the temperature above which the magnetic
order vanishes. Apart from the ordered and disordered phases,
we can identify the third intermediate one, which is also disordered but
exhibits a diverging susceptibility. Therefore, we specify one more transition
temperature denoted as $T_s$, above which the susceptibility divergence
disappears and the normal disordered phase begins.

\begin{figure}
\includegraphics[width=0.45\textwidth]{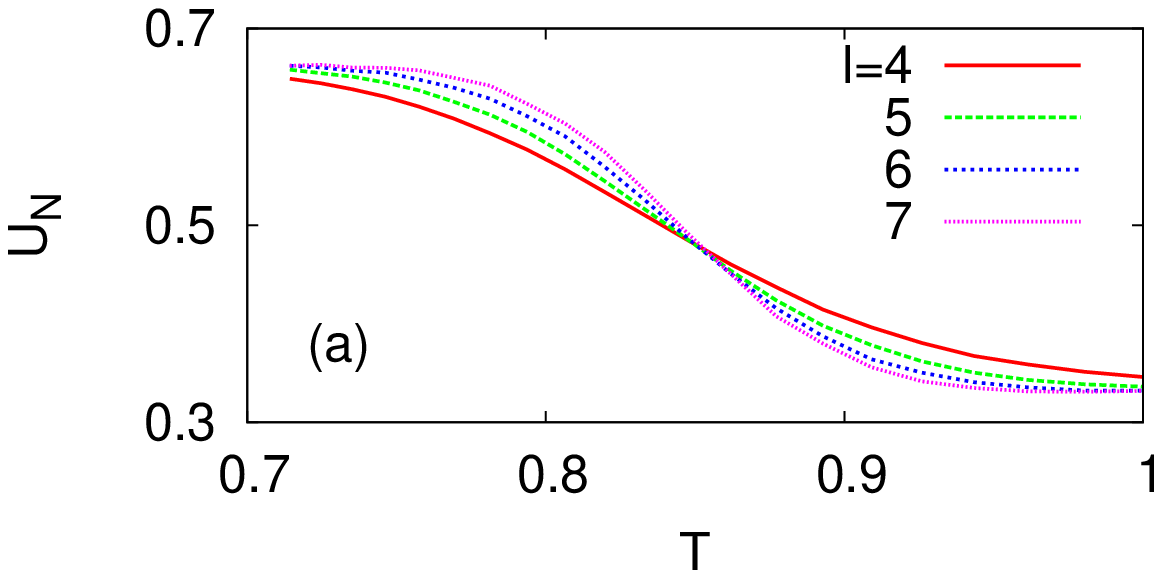}
\includegraphics[width=0.45\textwidth]{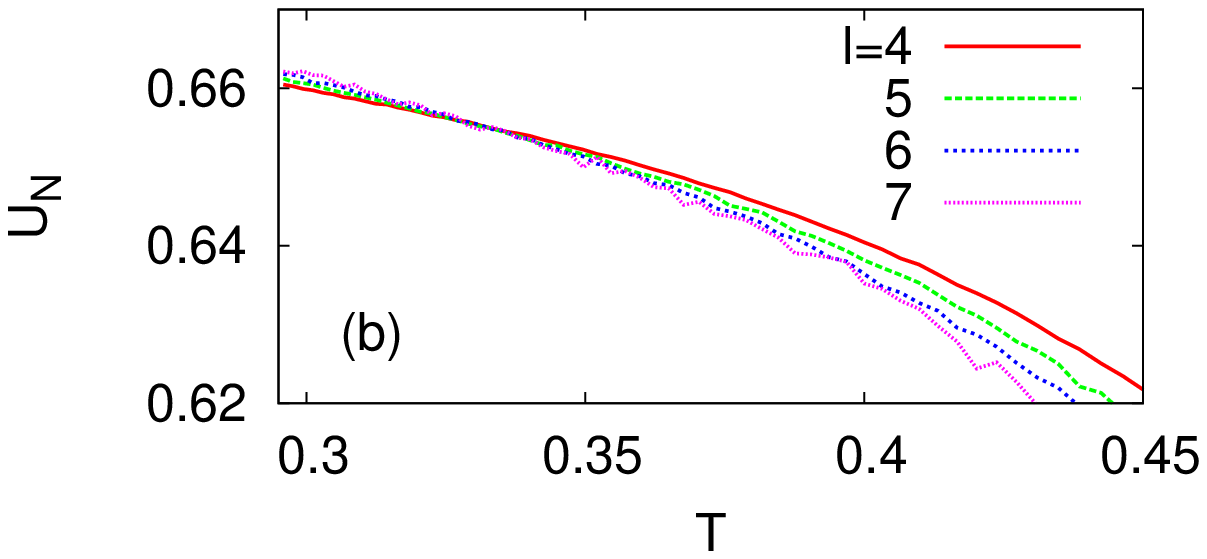}
\includegraphics[width=0.45\textwidth]{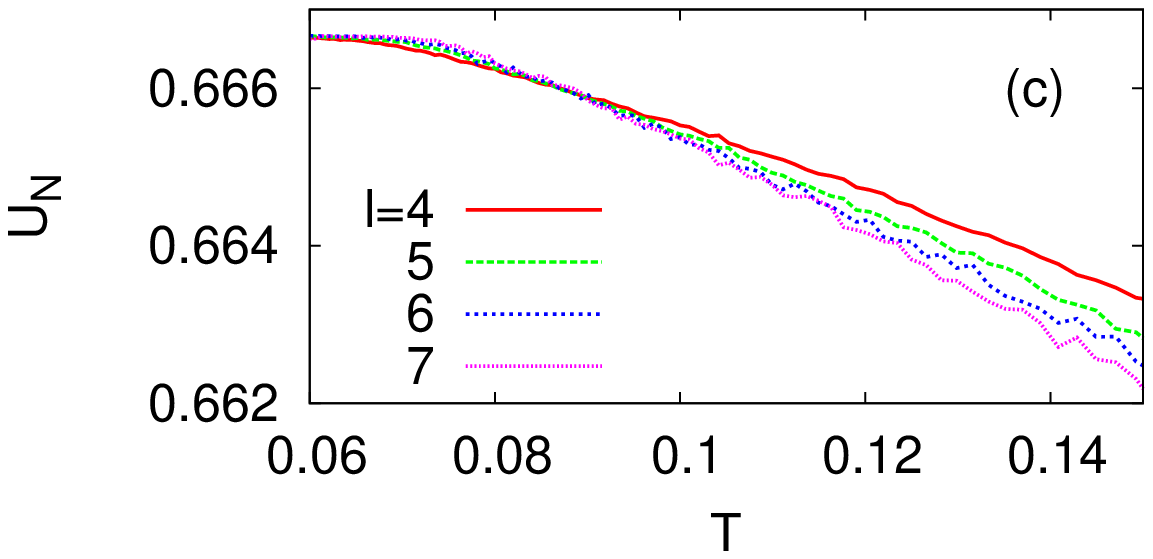}
\includegraphics[width=0.45\textwidth]{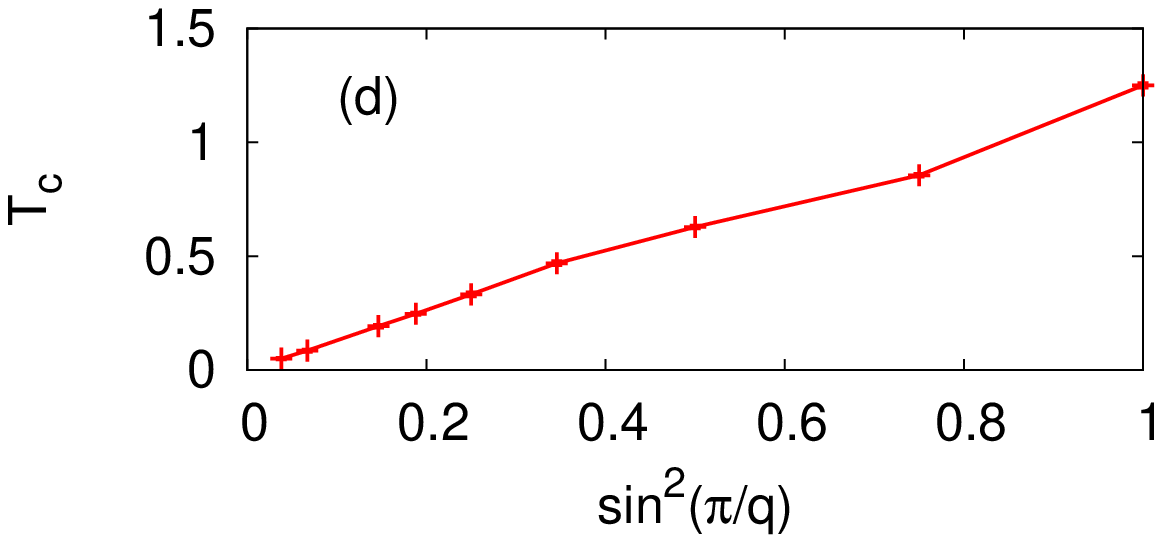}
\caption{(Color online) Binder's fourth-order cumulant for the $q$-state
clock model on heptagonal lattices with (a) $q=3$, (b) $q=6$, and (c)
$q=12$.
(d) Transition temperatures between the ordered and disordered phases, $T_c$,
compared to $\sin^2(\pi/q)$.}
\label{fig:binder}
\end{figure}

In order to obtain the phase diagram,
we employ the parallel tempering method~\cite{exchange} and measure the
magnetic order parameter
\[
\left< |m| \right>  = \left< \left| \frac{1}{N} \sum_j e^{i \theta_j}
\right| \right>,
\]
where $\left< \cdots \right>$ represents the thermal average.
From Binder's fourth-order cumulant~\cite{binder},
\[
U_N(T) = 1 - \frac{\left< |m|^4 \right>}{3 \left< |m|^2 \right>^2}
\]
for different $l$, we can locate a unique crossing point for each $q$
[Figs.~\ref{fig:binder}(a)-\ref{fig:binder}(c)].
This determines the lower transition temperature $T_c$ as a function of $q$.

Figure~\ref{fig:binder}(d) shows the dependence of $T_c$ on $q$.
$T_c$ is found to 
rapidly decrease to zero as $q$ grows larger. 
A striking observation is that $T_c$ is determined by
the typical energy scale $\Delta E$ to rotate a spin
in the fully ordered ground state. 
$\Delta E$
is roughly given by
\begin{equation}
\Delta E \propto 1 - \cos \left( \frac{2 \pi}{q} \right) = \sin^2 \left(
\frac{\pi}{q} \right)
\label{eq:sin2}
\end{equation}
in units of $J/k_B$ [see Eq.~(\ref{eq:hamiltonian})],
being
proportional to $T_c$ for each $q$
as clearly shown in Fig.~\ref{fig:binder}(d).
In addition, 
Eq.~(\ref{eq:sin2}) can be approximated by $T_c \propto
1/q^2$ for large $q$,
which is analogous to the planar case.
More interesting is the fact that the relation of $T_c \propto \Delta E$
captures the exact relation,
$T_c(q=4) = \frac{1}{2} T_c(q=2)$, mentioned in the previous section.
These results are 
consistent with the interpretation that the spin-wave
excitation breaks every magnetic order in the $XY$ model~\cite{xy};
in fact, Eq.~(\ref{eq:sin2}) leads to $\Delta E = 0$ in the limit of $q\to\infty$,
and thus $T_c=0$.

It is worthy to mention the significant contribution
of boundary spins to the determination of $T_c$; this is 
caused by the fact that the actual magnitude of 
$\Delta E$ depends on the number of neighbors.
Since boundary spins have fewer neighbors, 
the proportionality constant in Eq.~(\ref{eq:sin2})
takes a smaller value than those of bulk spins
so that their orientation will be strongly disturbed by thermal
fluctuations.
We also comment that the spin-wave excitation is observable
in the hyperbolic lattice without boundary since it is 
the basic excitation mode. 
In the latter system, however, the excitation is not 
sufficient to destroy the ordered phase
but arises as a separate peak in specific heat at $q>4$~\cite{gendiar}.

\begin{figure}
\includegraphics[width=0.45\textwidth]{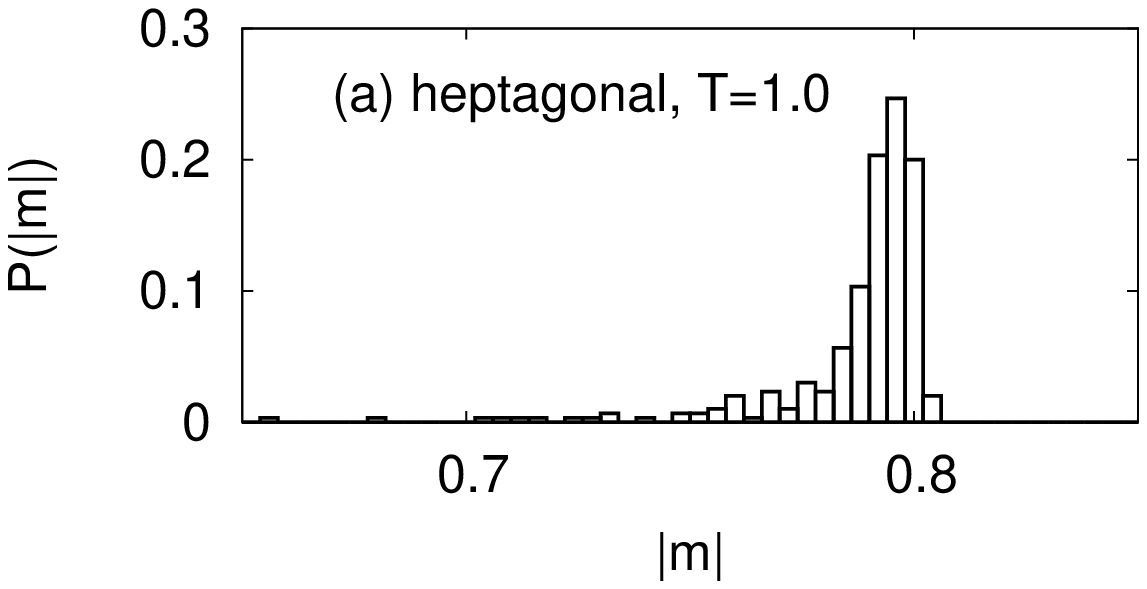}
\includegraphics[width=0.45\textwidth]{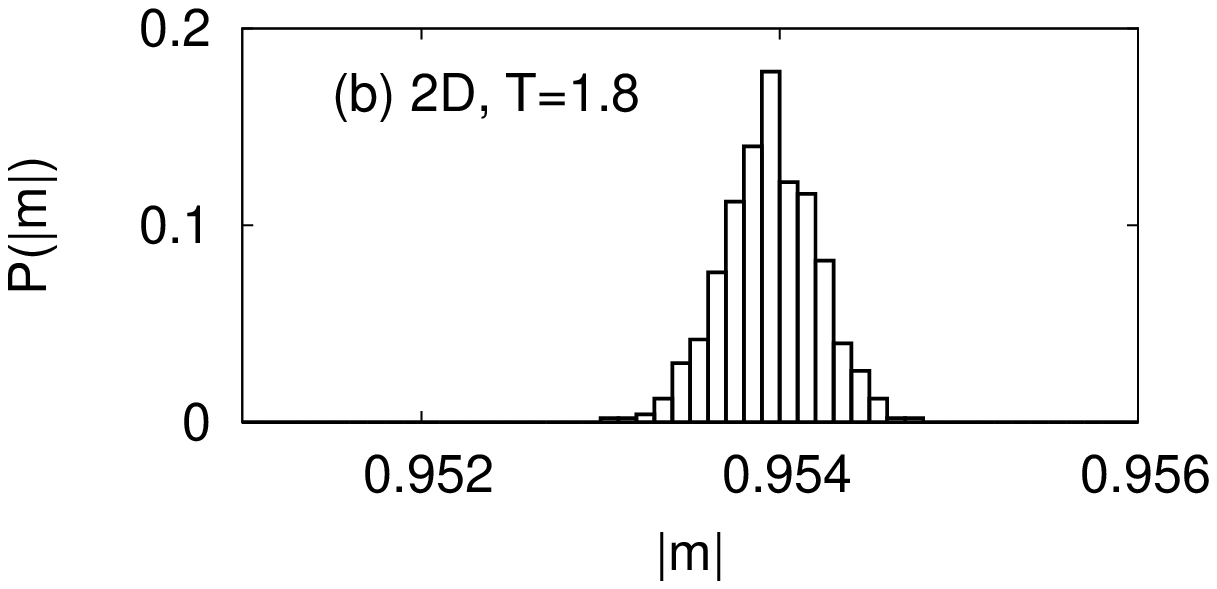}
\caption{Distributions of the magnetic order parameter
for Ising spin systems ($q=2$) (a) in a heptagonal lattice and (b) in a
plane. To compare these two cases,
we make both systems have sizes of $N \sim O(10^3)$,
and set $T \approx 0.8~ T_c$ to observe low-temperature regions.
A narrow Gaussian peak is clearly shown in the planar case
while a longer tail is
observed at low $|m|$ in the heptagonal case~\cite{auriac}.}
\label{fig:hist}
\end{figure}

\begin{figure}
\includegraphics[width=0.45\textwidth]{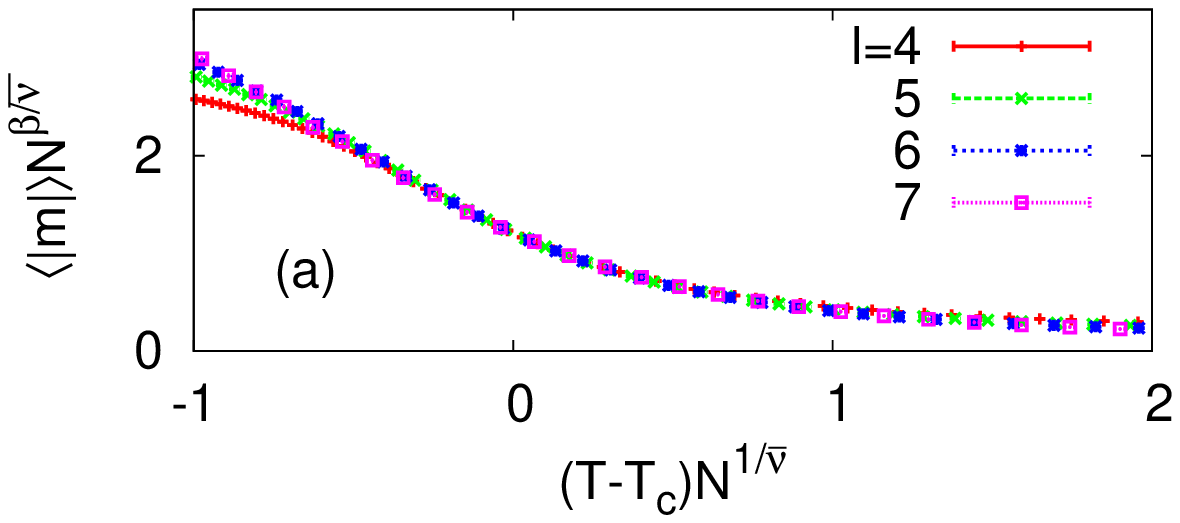}
\includegraphics[width=0.45\textwidth]{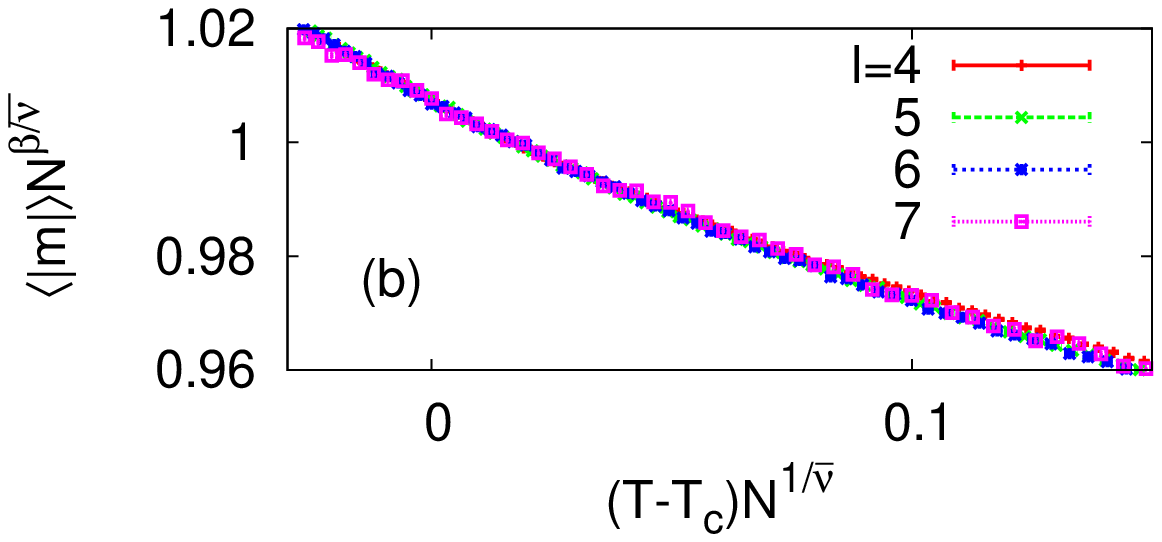}
\includegraphics[width=0.45\textwidth]{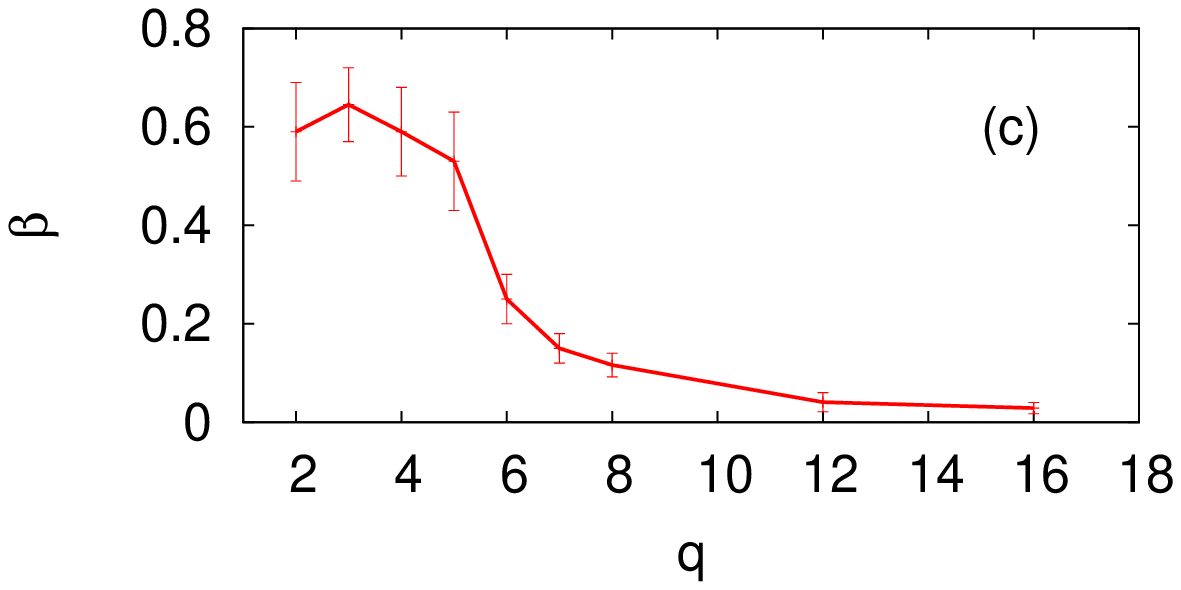}
\includegraphics[width=0.45\textwidth]{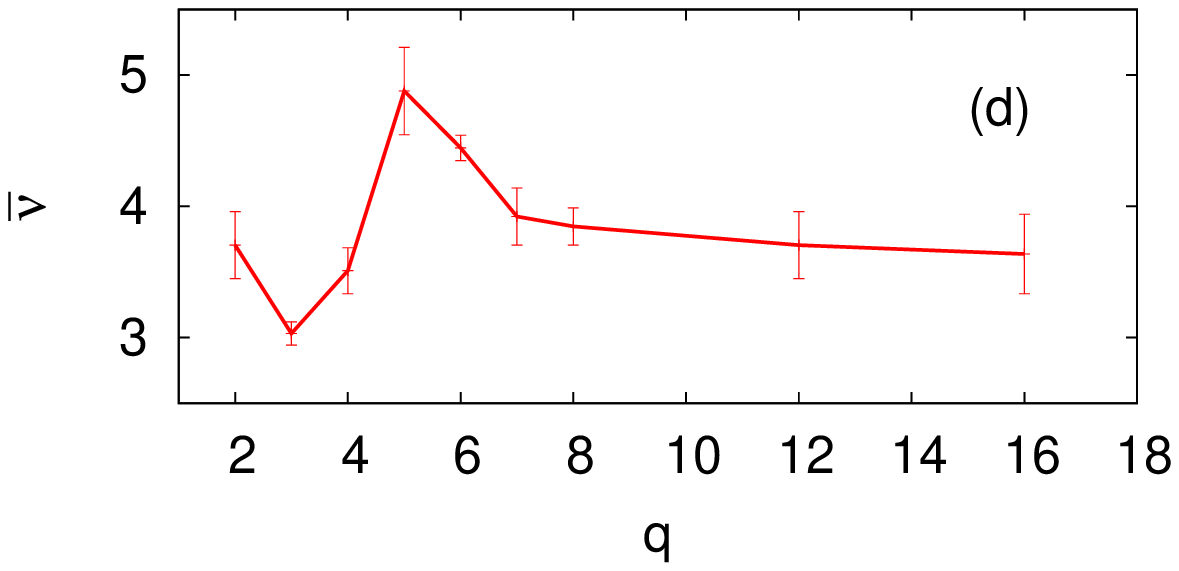}
\caption{(Color online)
Finite-size scaling using Eq.~(\ref{eq:fss}) for (a) $q=4$ and (b)
$q=16$. Here the parameters are chosen as $\beta/\bar\nu=0.167$,
$1/\bar\nu = 0.285$, and $T_c=0.63$ for $q=4$, and $\beta/\bar\nu=0.0075$,
$1/\bar\nu = 0.275$, and $T_c=0.05$ for $q=16$.
 By this way, we estimate behaviors of critical indices (c) $\beta$
and (d) $\bar\nu$ as $q$ varies.}
\label{fig:beta-nu}
\end{figure}

\subsection{Two distinct scaling relations around $T_c$ and $T_s$}
\label{???}

We next evaluate critical exponents of the transition
by employing the finite-size scaling analysis.
As pointed out in Ref.~\cite{auriac}, 
distribution functions of $|m|$
for the heptagonal lattice deviate from the Gaussian distributions (Fig.~\ref{fig:hist}).
Since the idea of the fourth-order cumulant assumes a Gaussian peak
shape~\cite{binder}, a direct scaling of the cumulant will give a
different value from the actual correlation-length exponent
estimated from the order parameter~\cite{shima}. In order to find an
appropriate estimate, therefore, we
perform at  each $q$ the scaling analysis
for $\left<|m|\right>$
based on the
scaling hypothesis:
\begin{equation}
\left< |m| \right> \propto N^{-\beta/\bar\nu} f\left( |T-T_c| N^{1/\bar\nu}
\right).
\label{eq:fss}
\end{equation}
In the present case, we choose $N$
instead of $l$
as a proper scaling variable, as $N$ gives
much better scaling collapse at $T_c$ than $l$.
Although the finite-size scaling of the Binder's cumulant with $\bar\nu$
fails due to the
non-Gaussian nature of the magnetization distribution, $T_c$'s estimated
from the crossing of $U_N$ and from Eq.~(\ref{eq:fss}) are almost identical.
Figure \ref{fig:beta-nu} shows
the resulting scaling plots and estimated critical exponents
as functions of $q$.
While $\bar\nu$ appears to be relatively
constant at $q>6$, $\beta$ tends to decrease to zero,
suggesting that that every $q$-state clock model belongs to a different
universality class, apart from the exact equivalence between $q=2$ and $4$.

\begin{figure}
\includegraphics[width=0.45\textwidth]{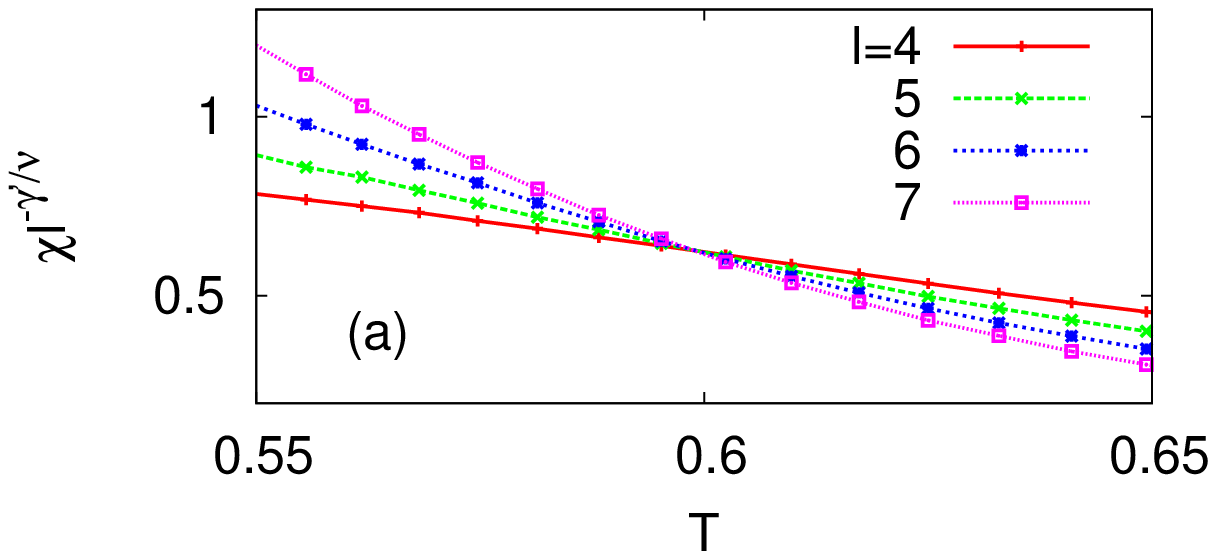}
\includegraphics[width=0.45\textwidth]{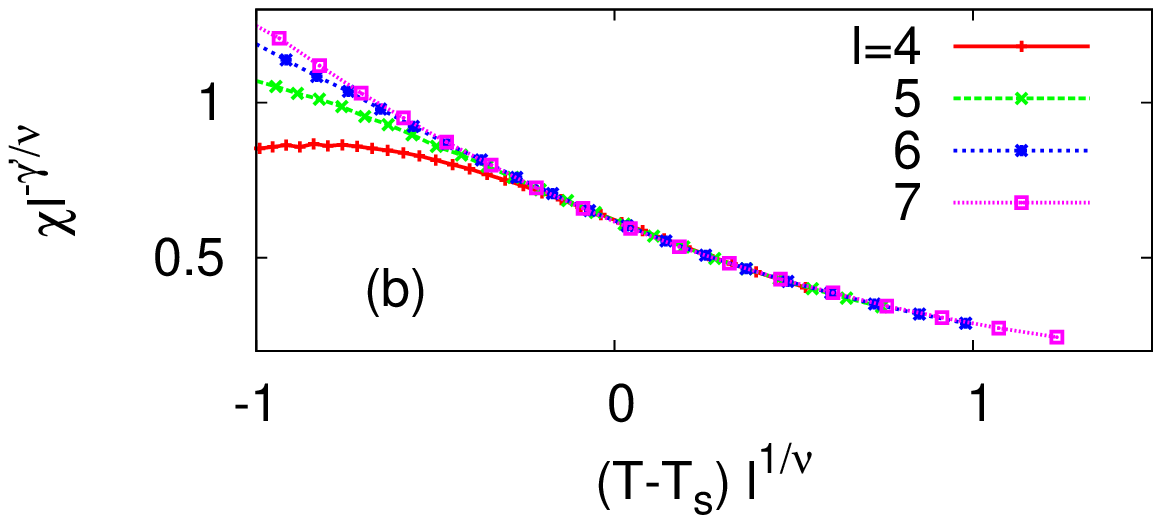}
\caption{(Color online) Susceptibility scaling with $l$ in
case of $q=6$. (a) The crossing point at $T_s=0.6$ with $\gamma'/\nu = 2$,
and (b) scaling collapse with $1/\nu = 1.5$.}
\label{fig:scalel}
\end{figure}

Measuring the magnetic susceptibility
$\chi = N (\left< |m|^2 \right> - \left< |m| \right>^2) /k_B T$
usually gives another way to estimate $\bar\nu$ with a similar scaling
hypothesis,
\begin{equation}
\chi \propto N^{-\gamma/\bar\nu} f\left( |T-T_c| N^{1/\bar\nu}
\right).
\label{eq:fss2}
\end{equation}
This yields consistent results with the above
ones for $q \le 4$, and confirms the results in Ref.~\cite{shima} for
$q=2$ and $4$.
However, we find Eq.~(\ref{eq:fss2}) inapplicable at $q > 4$ to
obtain critical indices since the susceptibility begins to diverge at a
temperature $T_s$, much higher than $T_c$. In contrast, the length scale
$l$ successfully works as a scaling variable (Fig.~\ref{fig:scalel}).
Henceforth, we should employ the following alternative scaling hypothesis
around $T_s$,
\begin{equation}
\chi \propto l^{-\gamma'/\nu} g(|T-T_s|~ l^{1/\nu}),
\label{eq:fss3}
\end{equation}
which locates the phase-separation point $T_s$ as depicted in
Fig.~\ref{fig:phase}.
In a usual $d$-dimensional lattice, there exists a trivial relationship
between exponents found in Eqs.~(\ref{eq:fss2}) and (\ref{eq:fss3}), derived
from $N \sim l^d$. In absence of such a relation between $N$ and $l$,
it is rather nontrivial to observe these different scalings in a
single system at different temperatures.
A similar change in the scaling variable across two transitions is also found
in percolation phenomena on hyperbolic lattices~\cite{perc}.

It is noticeable that a diverging susceptibility at finite $T_s$
appears to be the counterpart of the susceptibility divergence at $T_{\rm KT}$
in the planar $XY$ model.
More interestingly,
the higher transition temperature $T_s$, separate from $T_c$,
exists for all $q$ for the heptagonal lattice,
whereas
the quasiliquid phase 
in the planar case
does not appear with $q \le 4$.
To look into its
origin, we below examine the clock model on the Cayley tree.

\subsection{Comparison with Cayley tree}
\label{sub:cayley}

The Cayley tree is a special type of hyperbolic lattices containing no
loops, which often allows exact calculations as a useful guidance.
In order to understand the existence of the intermediate phase,
we extend the results for the Ising model ($q=2$)
on the Cayley tree, presented in Refs.~\cite{stosic1} and \cite{stosic2},
to general $q$.

Let us consider a branching number of $B=2$, i.e., a binary tree
with $n$ generations, where a root node is denoted as the zeroth generation.
The
total number of
nodes are $N_n = 2^{n+1}-1$. Let $Z_n^{(\theta)}$ denote the partition
function of this branch, restricted that the root node has a phase variable
as $\theta$. The complete partition function would be then $Z_n =
\sum_{\theta} Z_n^{(\theta)}$, where the summation $\sum_{\theta}$ runs over
$\theta = 0, \frac{2\pi}{q}, \frac{4\pi}{q}, \ldots, \frac{2\pi (q-1)}{q}$.
A tree with $(n+1)$ generations can be generated by attaching two trees with
$n$ generations to a single node, which has a certain angle $\theta$.
Since two $n$-generation trees are totally independent,
one can write the following recursion relation,
\begin{equation}
Z_{n+1}^{(\theta)} = e^{\bar\beta h \cos\theta} \left[ \sum_{\theta'}
Z_n^{(\theta')} e^{\bar\beta J \cos(\theta - \theta')} \right]^2,
\label{eq:recur}
\end{equation}
where $\bar\beta \equiv (k_B T)^{-1}$ and the magnetic field $h$ is assumed to
be in parallel with $\theta=0$.

By differentiating Eq.~(\ref{eq:recur}) and
taking the limit as $h \rightarrow 0$ (see Appendix~\ref{app:mag}),
we find the magnetization
of the $n$-generation tree with broken symmetry as follows:
\begin{eqnarray}
\left< m \right>_n &=& \frac{1}{N_n}
\frac{1}{Z_{n}^{(0)}}
\frac{\partial Z_{n}^{(0)}}{\partial(\bar\beta h)} =
\frac{1}{N_n} \sum_{j=0}^{n} (2R)^j \nonumber\\
&=& \frac{1}{2^{n+1}-1} \frac{(2R)^{n+1} - 1}{2R-1}, \label{eq:mag}
\end{eqnarray}
where $R \equiv \left[ \sum_{\theta} e^{\bar\beta J \cos\theta} \cos\theta
\right] / \left[\sum_{\theta} e^{\bar\beta J \cos\theta} \right]$.
Following the argument in Ref.~\cite{falk}, we remark that
the correlation between a pair of spins, separated by the distance
$r$, is given as $R^r$. Since
$R<1$ in general, the magnetization $\left< m \right>_n$ goes to zero as $n
\rightarrow \infty$.
One may also consider the free-energy cost of
forming a spin cluster on a subbranch of this tree. Since a single bond
divides the whole tree into two regions, the energy cost at the interface
is $\delta E = 2 \left[ 1-\cos(2\pi/q) \right]$, basically constant
regardless of the cluster size.
At any finite temperatures, the entropy gain $\delta S$, by forming a spin
cluster, will thus dominate the free-energy change, readily breaking the
magnetic order~(see Ref.~\cite{melin} for a typical spin configuration).
Yet one should note that this large-system limit can be quite
subtle~\cite{stosic3}.

The second-order derivative of Eq.~(\ref{eq:recur}) leads to (see
Appendix~\ref{app:sus})
\begin{widetext}
\begin{eqnarray}
\chi_n &=& \frac{\bar\beta}{N_n} \left[ \frac{1}{Z_n} \frac{\partial^2 Z_n}{\partial
(\bar\beta h)^2} - \left( \frac{1}{Z_n} \frac{\partial Z_n}{\partial (\bar\beta h)}
\right)^2 \right]
\simeq \frac{\bar\beta}{N_n q} \sum_{\theta} \frac{1}{Z_n^{(\theta)}}
\frac{\partial^2 Z_n^{(\theta)}}{\partial
(\bar\beta h)^2} \label{eq:sus}\\
&=& \frac{\bar\beta S}{(2^{n+1}-1)q} \left\{ \frac{(R+1)^2 2^{n+1}}{1-2R^2}
+ \frac{4R^4}{2R^2-1} \left[\sum_{j=0}^{n-1} (2R)^j \right]^2 +
\frac{2R+1}{2R^2-1} \sum_{j=0}^{n+1} (2R)^j \right\}\nonumber,
\end{eqnarray}
\end{widetext}
where $S \equiv \sum_\theta \cos^2 \theta$.
This formula recovers the Ising case with
$q=2$, where $\chi_n$ diverges at $\bar\beta_s J \equiv J/k_B T_s =
\ln(1+\sqrt{2}) \approx 0.8814$~\cite{stosic2}.
Also in general, Eq.~(\ref{eq:sus}) diverges at $R = 1/\sqrt{2}$.
If $q \rightarrow \infty$, we may rewrite the summations in $R$ as integrals
so that $R = \left[ \int_0^{2\pi} e^{\bar\beta J \cos\theta} \cos\theta \right]/
\left[ \int_0^{2\pi} e^{\bar\beta J \cos\theta} \right] = I_1(\bar\beta J) / I_0
(\bar\beta J)$,
where $I_n(x)$ is the modified Bessel function of the first kind.
A numerical solution then gives $\bar\beta_s J \approx 2.0582$. Since $\chi_n
(R)$ is basically the same at any $q$ and $R$ is always a monotonic function
of temperature, the divergence should be also qualitatively the same as in
$q=2$. That is, susceptibility diverges as $\chi_n / n \sim
a_0 + a_1 (T-T_s) n$ with some constants $a_0$ and $a_1$~\cite{stosic2}.

The susceptibility divergence can be explained by the presence of
boundaries~\cite{falk}.
From the viewpoint of boundary spins, which dominate the overall property,
the effective number of generations appears as $n' \simeq 2n$ since it is
the maximum possible distance in this tree. Therefore,
the effective branching number for a boundary spin
amounts to $\tilde{B} \simeq \sqrt{2}$
so that $\tilde{B}^{n'} = N \sim 2^n$ (see Ref.~\cite{auriac} for a
general discussion).
According to Eq.~(\ref{eq:sum}), the contribution of each boundary spin to
susceptibility is roughly $C = \sum_j^{n'} \tilde{B}^j R^j$.
Since the number of boundary spins is
proportional to the system size $N$, we find the lower bound of
susceptibility that $\chi_n \ge \bar\beta C$, and expectedly
this will make the most dominant term.
At $R = 1/\sqrt{2}$, we have $\tilde{B} R \simeq 1$,
which means that $C \simeq \sum^{2n} O(1)$. Note that the summation is
limited by the number of generations, $n$.
In other words, the susceptibility diverges with $\chi_n \propto n$ at
$R=1/\sqrt{2}$.

Recalling differences between with and without loops in percolation
phenomena~\cite{perc},
we may expect only a qualitative understanding for the heptagonal lattice
from studying the Cayley tree
rather than a quantitative agreement.
Although the presence of closed loops will presumably alter
the results described above,
the essential parts of these arguments could be conveyed to our heptagonal
lattice. That is, the susceptibility divergence at $T_s$ should be
attributed to the exponential growth of $N(l)$.
This is markedly different
from the case in regular lattices, where the susceptibility divergence is
due to divergence in the correlation length.
In particular, the
correlations among boundary spins play the most important role at this
point. Nonetheless, the correlation function does not have to decay
algebraically yet, which is a possible reason that Binder's cumulant does not
detect $T_s$. One cannot observe the algebraic decay until reaching $T_c
(< T_s)$. Around that point, the hyperbolic lattice begins to manifest itself
more as a surface. In contrary to the tree case above, for example,
the energy cost at a domain wall increases roughly
logarithmically with the cluster size~\cite{perc}, opening the possibility
for $T_c$ to be finite. As a consequence, we observe these three phases in
general: an ordered phase, a disordered phase but having a
diverging susceptibility, and a normal disordered phase with a finite
susceptibility.

\section{Summary}
\label{sec:summary}

We investigated the $q$-state clock model on the heptagonal lattice, and
found that the spin-wave excitation is relevant in the order-disorder
transition in this system. In the planar $q$-state clock model,
one could expect one additional
quasiliquid phase, and thus two phase transitions for $q>4$. The lower
transition defines the line between true- and quasi-long-range order, and
the higher one defines where the quasi-long-range order vanishes.
If we only introduce the curvature effect but without the finite
surface-volume ratio, the quasi-long-range order becomes a genuine order
and the higher transition is of the mean-field type
since fluctuation decays exponentially (see Sec~\ref{sub:hom}).
However, the presence of a boundary cannot be neglected,
which breaks the mean-field picture, and the spin-wave excitation appears to
be crucial in establishing the ordered phase.
In the limit of $q \rightarrow \infty$, the excitation
becomes gapless so that the transition temperature approaches zero.
In addition, the susceptibility begins to diverge at a higher temperature,
indicating a similar phenomenon to the KT transition with a diverging
susceptibility. By analyzing the clock model on the Cayley tree,
we suggest that the hyperbolic nature of the underlying lattice structure
makes the third phase observable for every $q \ge 2$.

\acknowledgments
S.K.B. and P.M. acknowledge the support from the Swedish Research Council
with the Grant No. 621-2002-4135, and B.J.K. is supported by the Korea
Science and Engineering Foundation through Grant No. R01-2007-000-20084-0.
H.S. is thankful for the supports by
a Grant-in-Aid for Scientific Research from Japan Society
for the Promotion of Science (Contract No.~19360042).
This research was conducted using the resources of High Performance
Computing Center North (HPC2N).

\begin{widetext}
\appendix

\section{Magnetization in Cayley Tree}
\label{app:mag}

If we take the limit of $h \rightarrow 0$, Eq.~(\ref{eq:recur}) leads to
\begin{equation}
Z_{n+1}^{(\theta)} =  {Z_n^{(\theta)}}^2 \left[ \sum_{\theta'} e^{\bar\beta J
\cos(\theta - \theta')} \right]^2,
\label{eq:partition}
\end{equation}
since $Z_n^{(\theta)} = Z_n^{(\theta=0)}$ by symmetry.
It is straightforward to see that
\begin{equation}
Z_n^{(\theta)} = \left[ \sum_{\theta'} e^{\bar\beta J \cos\theta'}
\right]^{2 N_n},
\label{eq:partition2}
\end{equation}
since
\begin{equation}
\sum_{\theta'} e^{\bar\beta J \cos(\theta - \theta')}
= \sum_{\theta'} e^{\bar\beta J \cos \theta'}.
\label{eq:identity1}
\end{equation}
Note that Eq.~(\ref{eq:partition2}) is an analytic function at any
$T$~\cite{egg}.
As to derivatives, one finds the following equations by differentiating
Eq.~(\ref{eq:recur}) with respect to $\bar\beta h$:
\begin{equation}
\frac{\partial Z_{n+1}^{(\theta)}}{\partial(\bar\beta h)} =
Z_{n+1}^{(\theta)} \cos \theta +
2 e^{\bar\beta h \cos\theta}
I_0 I_1,
\label{eq:recur1}
\end{equation}
where
$$
I_k (\theta) \equiv \sum_{\theta'} \frac{\partial^k Z_n^{(\theta')}}{\partial (\bar\beta h)^k}
e^{\bar\beta J \cos(\theta - \theta')}, \quad [k=0,1,2].
$$
We then take the zero-field limit, $h \rightarrow 0$.
By mathematical induction (see Appendix~\ref{app:mi}), one can show
\begin{equation}
\frac{\partial Z_n^{(\theta)}}{\partial (\bar\beta h)} = 
\frac{\partial Z_n^{(\theta=0)}}{\partial (\bar\beta h)} \cos \theta.
\label{eq:cos}
\end{equation}
Henceforth, by Eqs.~(\ref{eq:partition}) and (\ref{eq:cos}), we can rewrite
Eq.~(\ref{eq:recur1}) for a restricted ensemble with $\theta=0$ as follows,
\begin{equation}
\frac{1}{Z_{n+1}^{(0)}}
\frac{\partial Z_{n+1}^{(0)}}{\partial(\bar\beta h)} =
1 + 2 \frac{\sum_{\theta} e^{\bar\beta J \cos\theta} \cos\theta}
{\sum_{\theta} e^{\bar\beta J \cos\theta}}
\frac{1}{Z_n^{(0)}} \frac{\partial Z_n^{(0)}}{\partial
(\bar\beta h)},
\label{eq:restrict}
\end{equation}
which directly leads to Eq.~(\ref{eq:mag}).

\section{Mathematical Induction}
\label{app:mi}

Let us assume that Eq.~(\ref{eq:cos}) holds true, as it does for $n=0$,
\[
\left. \frac{\partial Z_0^{(\theta)}}{\partial (\bar\beta h)} \right|_{h=0} =
\left. e^{\bar\beta h \cos \theta} \cos \theta \right|_{h=0} = \cos \theta.
\]
Then for general $n$, this assumption yields the following relation:
\begin{equation}
\frac{\partial Z_{n+1}^{(\theta)}}{\partial(\bar\beta h)} =
Z_{n+1}^{(\theta)} \cos \theta + 2 Z_n^{(\theta)}
\frac{\partial Z_n^{(\theta=0)}}{\partial (\bar\beta h)}
\left[ \sum_{\theta'} e^{\bar\beta J \cos(\theta - \theta')} \right]
\left[ \sum_{\theta'}
e^{\bar\beta J \cos(\theta - \theta')} \cos \theta' \right].
\label{eq:ind}
\end{equation}
Here we note the following identity:
\begin{eqnarray}
\sum_{\theta'} e^{\bar\beta J \cos(\theta - \theta')} \cos \theta'
&=& \sum_{\theta''} e^{\bar\beta J \cos\theta''} \cos (\theta -
\theta'')\nonumber\\
&=& \cos \theta \sum_{\theta''} e^{\bar\beta J \cos\theta''} \cos \theta''
+ \sin \theta \sum_{\theta''} e^{\bar\beta J \cos\theta''} \sin
\theta''\nonumber\\
&=& \cos \theta \sum_{\theta''} e^{\bar\beta J \cos\theta''} \cos \theta'',
\label{eq:identity2}
\end{eqnarray}
where $\theta'' \equiv \theta - \theta'$ and the last equality is due to
the fact that $\sin \theta''$ is an odd function. Therefore, we substitute
Eqs.~(\ref{eq:identity1}) and (\ref{eq:identity2}) into Eq.~(\ref{eq:ind})
and then obtain
\begin{eqnarray*}
\frac{\partial Z_{n+1}^{(\theta)}}{\partial(\bar\beta h)} &=&
\left\{ Z_{n+1}^{(\theta=0)} + 2 Z_n^{(\theta=0)}
\frac{\partial Z_n^{(\theta=0)}}{\partial (\bar\beta h)}
\left[ \sum_{\theta'} e^{\bar\beta J \cos \theta'} \right]
\left[ \sum_{\theta'}
e^{\bar\beta J \cos \theta'} \cos \theta' \right] \right\} \cos \theta\\
&=& \frac{\partial Z_{n+1}^{(\theta=0)}}{\partial(\bar\beta h)} \cos \theta,
\end{eqnarray*}
which confirms Eq.~(\ref{eq:cos}) for any $n \ge 0$.

\section{Susceptibility in Cayley Tree}
\label{app:sus}
For describing susceptibility, we again differentiate Eq.~(\ref{eq:recur1})
to get
\begin{eqnarray}
\frac{\partial^2 Z_{n+1}^{(\theta)}}{\partial(\bar\beta h)^2} &=&
\frac{\partial Z_{n+1}^{(\theta)}}{\partial (\bar\beta h)} \cos \theta +
2 e^{\bar\beta h \cos\theta} 
\left(
\cos \theta I_0 I_1 + I_1^2 + I_0 I_2 
\right).
\label{eq:recur2}
\end{eqnarray}
In the zero-field limit, we have the following:
\begin{eqnarray*}
\frac{1}{Z_{n+1}^{(\theta)}}
\frac{\partial^2 Z_{n+1}^{(\theta)}}{\partial(\bar\beta h)^2} &=&
(1+4R) \cos^2 \theta \frac{1}{Z_n^{(0)}} \frac{\partial
Z_n^{(0)}}{\partial (\bar\beta h)}
+ 2 R^2 \cos^2 \theta \left[ \frac{1}{Z_n^{(0)}} \frac{\partial
Z_n^{(0)}}{\partial (\bar\beta h)} \right]^2
+ 2 \frac{\sum_{\theta'} \frac{1}{Z_n^{(\theta')}} \frac{\partial^2
Z_n^{(\theta')}}{\partial (\bar\beta h)^2} e^{\bar\beta J \cos(\theta -
\theta')}}{\sum_{\theta'} e^{\bar\beta J \cos \theta'}}.
\end{eqnarray*}
To simplify the last term, we sum up both sides
over $\theta$ and find
\begin{equation}
\sum_{\theta} \frac{1}{Z_{n+1}^{(\theta)}}
\frac{\partial^2 Z_{n+1}^{(\theta)}}{\partial(\bar\beta h)^2} =
S \left\{ (1+4R) \frac{1}{Z_n^{(0)}} \frac{\partial
Z_n^{(0)}}{\partial (\bar\beta h)}
+ 2 R^2 \left[ \frac{1}{Z_n^{(0)}} \frac{\partial
Z_n^{(0)}}{\partial (\bar\beta h)} \right]^2 \right\}
+ 2 \sum_{\theta} \frac{1}{Z_n^{(\theta)}} \frac{\partial^2
Z_n^{(\theta)}}{\partial (\bar\beta h)^2},
\label{eq:susrecur}
\end{equation}
where $S \equiv \sum_\theta \cos^2 \theta$. Now Eq.~(\ref{eq:susrecur})
describes the full ensemble without breaking symmetry, which is valid above
criticality. The terms inside the curly brackets can be explicitly written by
using Eq.~(\ref{eq:mag}).
Solving this recursion relation with the first term as
\[
\sum_{\theta} \frac{1}{Z_0^{(\theta)}} \frac{\partial^2
Z_0^{(\theta)}}{\partial (\bar\beta h)^2} = \sum_{\theta} \cos^2 \theta = S,
\]
we obtain Eq.~(\ref{eq:sus}) as the susceptibility for the $n$-generation
tree.

\end{widetext}


\end{document}